\documentclass[10pt,prc,showpacs,superscriptaddress,twocolumn]{revtex4}
\usepackage{graphicx}
\begin{document}
\title{Elliptic Flow Analysis at RHIC with the Lee-Yang Zeroes Method in a Relativistic Transport Approach}
\author{Xianglei Zhu}
\affiliation{Frankfurt Institute for Advanced Studies (FIAS), Max-von-Laue-Str.~1, D-60438 Frankfurt am Main,
Germany} \affiliation{Institut f\"ur Theoretische Physik, Johann Wolfgang Goethe-Universit\"at,
Max-von-Laue-Str.~1, D-60438 Frankfurt am Main, Germany} \affiliation{Physics Department, Tsinghua University,
Beijing 100084, China}
\author{Marcus Bleicher}
\affiliation{Institut f\"ur Theoretische Physik, Johann Wolfgang Goethe-Universit\"at, Max-von-Laue-Str.~1,
D-60438 Frankfurt am Main, Germany}
\author{Horst St\"ocker}
\affiliation{Frankfurt Institute for Advanced Studies (FIAS), Max-von-Laue-Str.~1, D-60438 Frankfurt am Main,
Germany} \affiliation{Institut f\"ur Theoretische Physik, Johann Wolfgang Goethe-Universit\"at,
Max-von-Laue-Str.~1, D-60438 Frankfurt am Main, Germany}
\begin{abstract}
The Lee-Yang zeroes method is applied to study elliptic flow ($v_2$) in Au+Au collisions at $\sqrt{s}=200A$~GeV, with the UrQMD model. In this transport approach, the true event plane is known and both the
nonflow effects and event-by-event $v_2$ fluctuations exist.
Although the low resolutions prohibit the application of the method for most central and peripheral collisions, the integral and differential elliptic flow from the Lee-Yang zeroes method agrees with the exact $v_2$ values very well for semi-central collisions. 
\end{abstract}
\maketitle

Anisotropic flow \cite{Bass:1993ce,Hartnack:1994bs,Bass:1995pj,Sorge:1998mk}, or more specifically, elliptic flow ($v_2$), which is
the second Fourier harmonic \cite{Voloshin} in the transverse distribution of the
emitted particles, is expected to be sensitive to the early pressure
gradients and therefore to the equation of state (EOS) of the formed
fireball in heavy-ion collisions \cite{Voloshin,Ollitrault,Stoecker:1986ci,Teaney_Hydro,hydro,Bleicher:2000sx}. 
In addition, the
elliptic flow of high $p_T$ particles is related to jet
fragmentation and energy loss of the primordially produced hard
antiquark-quark pair when traveling through the hot QCD medium
\cite{Gyulassy_Jet}. 
Therefore, understanding the microscopic nature of the elliptic flow is of great importance for the study of the key questions of experimental and theoretical relativistic heavy ion collisions, the transport coefficients and the pressure, namely the equation of state.

To measure flow, experiments usually use the reaction plane method \cite{Poskanzer_Voloshin,Molitoris:1985df} or the equivalent two-particle correlation method \cite{Poskanzer_Voloshin,STAR_flow,STAR_PRC66}. However, these two-particle correlation-based methods suffer from correlations unrelated to the reaction plane.  These additional contributions are
usually dubbed nonflow effects \cite{Borghini:2000cm}, such as the overall transverse momentum conservation, small angle azimuthal correlations due to final state interactions, resonance decays, jet production \cite{Kovchegov:2002nf} and quantum correlations due to the HBT effect \cite{Dinh:1999mn}.  
In order to decrease the contribution of the nonflow effects to the flow measurements, many-particle
cumulant method was proposed \cite{Borghini}. In this method, usually 4- and 6-particle cumulants are used to estimate the collective flow. This method has been applied to the flow analysis at RHIC, SPS and SIS/GSI \cite{STAR_flow,STAR_PRC66,STAR_highpt,Adams:2003zg,Tang:2004vc,STAR_Voloshin,PHENIX_PRL94,SPS_flow,Bastid:2005ct}. And it has also been tested in the UrQMD model in \cite{Zhu:2005qa}. It was found that the two-particle correlation method is strongly affected by nonflow effects in the UrQMD model, such as jet production and resonance decays, while the 4- and 6-particle cumulants give the true $v_2$ values with better than $<5\%$ accuracy for the semi-central collisions at RHIC, even though this method is affected severely by the rather large $v_2$ fluctuations for the central and very peripheral collisions \cite{miller03}. More recently, the Lee-Yang zeroes method was developed to analyze the collective flow \cite{Bhalerao:2003yq,Bhalerao:2003xf,Borghini:2004ke}. This new method enables the extraction of the ``true" collective flow from the genuine correlation between a large number of particles. It is expected to give the cleanest values of the genuine collective flow. It is, in fact, also easier to apply to data than the cumulant method. This method has been applied in the flow analysis at SIS/GSI \cite{Bastid:2005ct}. Here it was found that the $v_1$ and $v_2$ values extracted from this new method do agree with those from the higher order cumulant method. 

In the present paper, the Lee-Yang zeroes method will be tested in the integral and differential $v_2$ analysis of UrQMD events at RHIC energy. The UrQMD model \cite{urqmd,urqmd2.1} is well suited for the test: it has been shown in \cite{Zhu:2005qa} that both the nonflow effects and the $v_2$ fluctuations \footnote{The source of $v_2$ fluctuations still deserves further investigation, see for example \cite{Mrowczynski:2005gw}}, which can affect the flow data, do exist naturally in the model. Most importantly, the true reaction plane angle $\Phi_R$ is known by construction in the UrQMD model. This allows for the direct calculation of the exact elliptic flow from its basic definition, i.e. $v_2=\langle \cos2(\phi-\Phi_R)\rangle$. Through the present analysis, it will be demonstrated that the Lee-Yang zeroes method works well for the $v_2$ measurements at RHIC energy.

\begin{figure}[h]
\centering
\includegraphics[totalheight=2.3in]{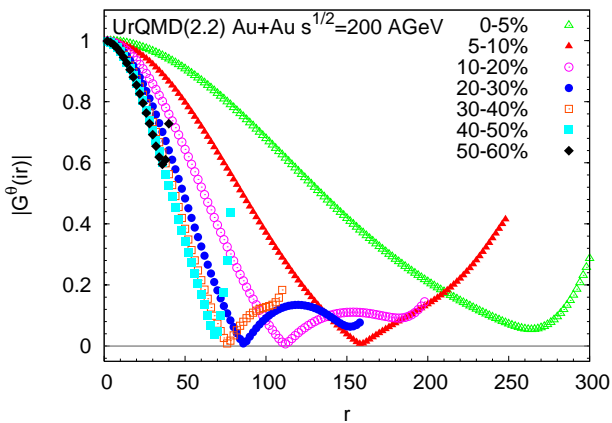}
\caption{(Color online) $|G^\theta(ir)|$ as a function of $r$ for $\theta=0$ at different centralities.} \label{fig1}
\end{figure}

In order to determine the ``integrated" elliptic flow, which is defined as 
$$V_2=\left\langle \sum_{j=1}^M \omega_j \cos(n(\phi_j-\Phi_R))\right\rangle,$$
the generating function of the Lee-Yang zeroes method is introduced and defined as \cite{Bhalerao:2003yq,Bhalerao:2003xf,Borghini:2004ke} $$G^\theta(ir)=\left\langle \prod_{j=1}^{M}\left[ 1+ir\omega_{j}\cos\left( \phi_{j}-\theta\right) \right]  \right\rangle.$$ In the above definitions, $M$ is the number of particles used to estimate the integral flow in each event, $\phi_j$ is the particle azimuthal angle, $r$ is a positive real variable, $\theta$ is an arbitrary reference angle and $\omega_{j}$ is the particle weight. It is worth stressing that in our analysis, the $1/M$ weight is used to decrease the effect of event-by-event $M$ fluctuations. Therefore, $V_2=v_2$ in our analysis. 
Fig. \ref{fig1} shows the amplitudes of the generating function $|G^{\theta}(ir)|$ as a function of $r$ for $\theta=0$ at different centralities. For the semi-central bins (e.g. 20-30\%), with the increase of $r$, $|G^{\theta}(ir)|$ decreases rapidly from the value of 1 at $r=0$, then reaches a sharp minimum which is very close to 0. This appearence of the sharp minimum indicates that there exists a zero of the generating function. After finding the position $r_0^{\theta}$ of the first zero of $|G^{\theta}(ir)|$, the integral elliptic flow is determined as
$$v_2^\theta=j_{01}/r_0^{\theta},$$
where $j_{01}=2.40483$ is the first root of the Bessel function $J_0(x)$. The final value of the integral $v_2$ is the average of $v_2^\theta$ over 5 equally spaced values of $\theta$ from 0 to $4\pi/5$, and the statistical uncertainty of the final integral $v_2$ is about a factor of 2 smaller \cite{Bhalerao:2003xf}.

\begin{figure}[h]
\centering
\includegraphics[totalheight=2.3in]{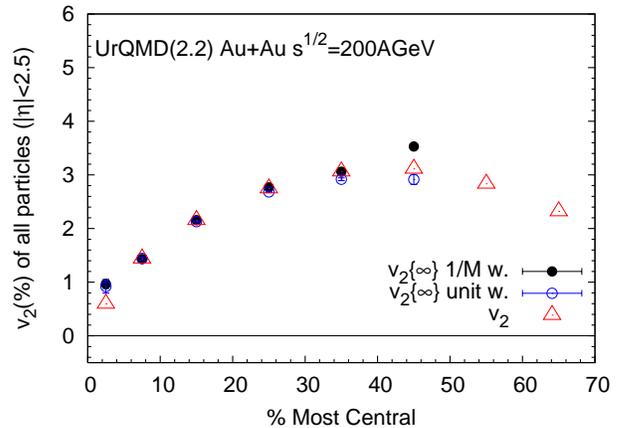}
\caption{(Color online) The UrQMD model results for integral $v_2$ values for Au+Au collisions at $\sqrt s=200A$~GeV. Results from the Lee-Yang zeroes method ($v_2\{\infty\}$) with unit weight and $1/M$ weight are compared to the exact $v_2$ in different centrality bins. (see text for details.) } \label{fig2}
\end{figure}

Fig. \ref{fig2} shows the Lee-Yang zeroes results on the centrality dependence of the integral $v_2$. 
For the integral $v_2$ analysis, all particles in the
pseudorapidity region $|\eta|<2.5$ are used and the number of particles from event to event fluctuates in each
centrality bin. The centralities in our
analysis are selected according to the same geometrical fractions of the total cross section
(0-5\%,5-10\%,10-20\%,20-30\%,30-40\%,40-50\%,50-60\%,60-70\%) as used by the STAR experiment \cite{STAR_flow}. Here,
however, we use impact parameter cuts instead of multiplicity cuts. More than $1.3\cdot 10^6$ minimum bias
events are used in the integral $v_2$ analysis.

\begin{figure}[h]
\centering
\includegraphics[totalheight=2.3in]{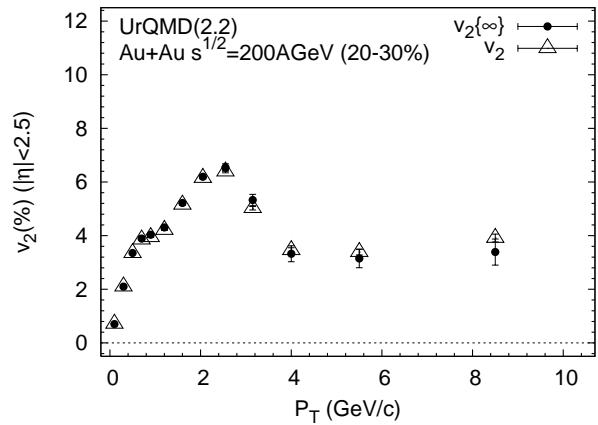}
\caption{The UrQMD model results for $v_2(p_T)$ in semi-central (20-30\%) Au+Au collisions at $\sqrt s=200A$~GeV. Results from the Lee-Yang zeroes method ($v_2\{\infty\}$) are compared to the exact $v_2$.} \label{fig3}
\end{figure}

The $v_2$ values extracted from the Lee-Yang zeroes method with $1/M$ weight (solid circles) agrees with the exact $v_2$ (open triangles) very well for the semi-central bins (about 5-40\%), as has been seen in Fig. \ref{fig2}. However, the Lee-Yang zeroes method does not work well for the most central (0-5\%) and peripheral ($>$40\%) bins in the present model study: the resolution parameter $\chi$, which is related to the event-plane resolution \cite{Ollitrault:1997vz}, decides whether the method can be applied or not \cite{Bhalerao:2003yq,Bhalerao:2003xf,Borghini:2004ke}. 
If this parameter is too small, i.e., $\chi \leq 0.5$, the statistical error will be too large, and the method is inapplicable.
$\chi$ is roughly given by $v_2\sqrt{M}$. Therefore, it will be smaller for the most central bin where $v_2$ is small, and for peripheral bins where $M$ is small. For the feasibility test with the UrQMD model, $\chi$ is about 1 in the semi-central bin (20-30\%), while it is close to 0.5 in the most central (0-5\%) and peripheral bins ($>$40\%). In fact, as is shown in Fig. \ref{fig1}, the first minimum of $|G^{\theta}(ir)|$ is not compatible with a zero for most central (0-5\%) and peripheral ($>$40\%) bins. Therefore, the Lee-Yang zeroes method can not be used in most central (0-5\%) and peripheral ($>$40\%) bins from the UrQMD model. 

We have also tried the Lee-Yang zeroes method with unit weight. These results are shown as open circles in Fig. \ref{fig2}. For the semi-central bins (10-40\%), it slightly undershoots the exact $v_2$, which indicates that the zeroes method with unit weight is not completely free from the effect of $M$ fluctuations. However, the effect of $M$ fluctuations is small. Experimentally, the centrality is usually selected according to the multiplicity $M$ of the events. In this case, the $M$ fluctuations will be narrower, and the effect of $M$ fluctuations on the zeroes method will be weaker. The present Lee-Yang zeroes analysis based on the $M$ selected centralities (not shown in this Letter) indeed tells that the integral $v_2$ values from the $1/M$ weight and unit weight agree well with each other and the $M$ fluctuations have almost negligible effects.

\begin{figure}[h]
\centering
\includegraphics[totalheight=2.3in]{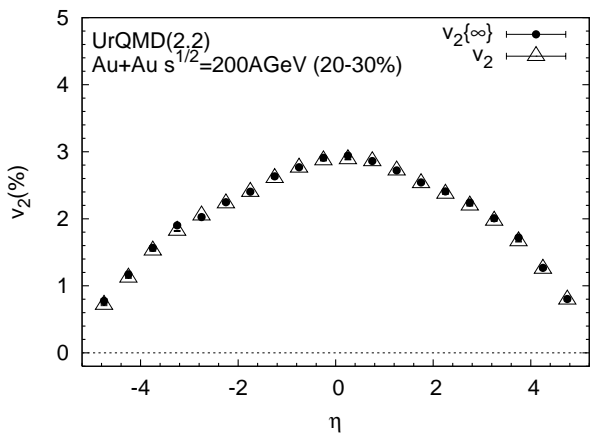}
\caption{The UrQMD model results for $v_2(\eta)$ in semi-central (20-30\%) Au+Au collisions at $\sqrt s=200A$~GeV. Results from the Lee-Yang zeroes method ($v_2\{\infty\}$) are compared to the exact $v_2$.} \label{fig4}
\end{figure}

After determining $r_0^{\theta}$, the differential $v_2$ can be estimated from Eq.~(9) in \cite{Borghini:2004ke}. For the test of the differential flow, we use more than $6\cdot 10^5$  semi-central
events (with impact parameters from 6.7 to 8.3~fm, corresponding to about 20\% to 30\% of the total cross
section). From the above results on the integral $v_2$, we know that the Lee-Yang zeroes method
reproduces the exact $v_2$ in this centrality bin with better than 2\% accuracy, However it is still necessary to see whether it reproduces the differential $v_2$ correctly. Especially at large transverse momenta ($p_T$), 
nonflow contributions are expected to be large. 
Fig. \ref{fig3} and Fig. \ref{fig4} shows the comparisons of the Lee-Yang zeroes results for $v_2(p_T)$ and $v_2(\eta)$ to the exact $v_2$ values.
It is clear that the results from the Lee-Yang zeroes method agree with the exact $v_2$ values very well for the whole $p_T$- and $\eta$-range. This shows that the Lee-Yang zeroes method indeed reproduced exactly the genuine elliptic flow values for these centrality bins very well.

To summarize, we have applied the Lee-Yang zeroes method in the $v_2$ analysis of the UrQMD model at 200$A$~GeV RHIC Au+Au collisions. Only for the semi-central bins (5-40\%), is this method actually applicable. Both the integral and differential $v_2$ from the Lee-Yang zeroes method agree well with the exact values in these bins. The UrQMD model includes many of the nonflow correlations which are also seen in the real data. Hence, these good agreements show that the Lee-Yang zeroes method is not sensitive to the nonflow correlations and $v_2$ fluctuations for the semi-central bins and indeed gives the genuine elliptic flow there. It should be noted that the current UrQMD model underpredict the RHIC $v_2$ data by about 40\%. Therefore, in the experiments at RHIC or LHC, the zeroes method will work over a somewhat broader range of centrality bins. 

\begin{acknowledgments}
This work was triggered by a discussion with Dr. A. Poskanzer at LBL. The authors thank Drs. A. Poskanzer, J.-Y. Ollitrault and R. Snellings for stimulating discussions. We are grateful to the Center for the Scientific Computing (CSC) at Frankfurt for the computing resources. This work was supported by GSI and BMBF. X.Z. thanks the Frankfurt International Graduate School for Science (FIGSS) at the J.~W.~Goethe-Universit\"at for financial support.
\end{acknowledgments}

\end{document}